\documentstyle[aps,prl,twocolumn,epsfig]{revtex} 
\draft
\begin{document}                                     
\pagestyle{myheadings}
\markboth{D. Helbing/M. Schreckenberg: Cellular Automata 
Simulating Experimental Properties of Traffic Flow}
{D. Helbing/M. Schreckenberg: Cellular Automata 
Simulating Experimental Properties of Traffic Flow}
\tighten
\onecolumn
\twocolumn[\hsize\textwidth\columnwidth\hsize\csname @twocolumnfalse\endcsname
{\protect
\title{Cellular Automata Simulating Experimental Properties of Traffic Flow}
\author{Dirk Helbing$^1$ and Michael Schreckenberg$^2$}
\address{$^1$II. Institute of Theoretical Physics, University of Stuttgart,
Pfaffenwaldring 57/III, 70550 Stuttgart, Germany\\
$^2$Theoretische Physik, Gerhard-Mercator-Universit\"at Duisburg,
Lotharstra{\ss}e 1, 47048 Duisburg, Germany}
\maketitle      
\begin{abstract}
A model for 1D traffic flow is developed, which is discrete in
space and time. Like the cellular automaton model by Nagel and Schreckenberg
[J. Phys. I France {\bf 2}, 2221 (1992)], it is simple, 
fast, and can describe stop-and-go traffic. Due to its relation
to the optimal velocity model by Bando {\em et al.} 
[Phys. Rev. E {\bf 51}, 1035 (1995)], its instability
mechanism is of deterministic nature. The model can be easily calibrated
to empirical data and displays the experimental 
features of traffic data recently reported
by Kerner and Reh\-born [Phys. Rev. E {\bf 53}, R1297 (1996)]. 
\end{abstract}
\pacs{05.50.+q,05.40.+j,47.55.-t,89.40.+k}
} ]
Cellular automata (CA) are interesting for their speed and their complex
dynamical behavior \cite{Wolfram}, including such fascinating 
phenomena as self-organized criticality \cite{SOC,aval,earth},
formation of spiral patterns \cite{Spiral}, or oscillatory and chaotic sequences of
states \cite{Wolfram,Spiral,chaos}. 
Their enormous computation speed
and efficiency is a consequence of the following properties, which
are ideal preconditions for parallel computing: 
(i) discretization of 
space into identical sites $x$, 
(ii) a finite number of possible states
$f(x)$ (iii) the (parallel) update at times $T = t\,\Delta T$
with an elementary time step $\Delta T$, (iv) globally applied update rules,
based on (v) short-range interactions
with a finite (small) number of neighbouring sites. 
Despite these simplifications, cellular automata have a broad range
of applications, reaching from realistic
simulations of granular media \cite{gran} or fluids \cite{fluid} 
(including interfacial phenomena and magnetohydrodynamics),
over the computation of chemical reactions \cite{Spiral,chem}, 
up to the modeling
of avalanches \cite{aval}.
Their application to traffic dynamics \cite{NS,Nag} 
has stimulated an enormous research activity \cite{traf,Hab,Letter},
aiming at an understanding and control of traffic instabilities, which
are responsible for stop-and-go traffic and congestion, both on `freeways' 
and in cities. 
\par
Recently, Kerner and Rehborn \cite{emp} have reported some characteristic
properties of empirical highway traffic flow, which a realistic traffic 
model should display: 
(i) At small densities, traffic flow is stable,
i.e., arbitrarily large disturbances of homogeneous
traffic will disappear in the course of time. 
(ii) Above a certain critical density, any small perturbation will give rise
to the formation of a traffic jam. 
(iii) Between the stable and the unstable regions, there
exists a density interval beginning at about 20 vehicles per kilometer,
where traffic flow is metastable. That is, 
sufficiently small disturbances (so-called `subcritical perturbations')
will fade away, whereas `supercritical' perturbations
exceeding a certain minimal amplitude will cause 
a traffic jam. 
(iv) The outflow from traffic jams has a typical value which 
is independent of the
initial conditions and, to a large extent, independent
of the average surrounding traffic density. It varies only 
with road or weather conditions, and the average vehicle characteristics
(regarding their lengths and acceleration capabilities). 
(v) The typical outflow 
is considerably smaller than the maximum flow and lies at about
$1600$\,vehicles/km for slow lanes, and on fast lanes
between $1800$\,vehicles/km (measured in Germany \cite{emp})
and $2100$\,vehicles/km (on Dutch motorways 
\cite{traf}). It is associated with
a density of $20\pm 5$ vehicles per kilometer.
(vi) Downstream jam fronts
move with a velocity of $-15\pm 5$\,km/h.
Notice that properties (iv) and (vi) originate from 
the uniform acceleration behavior of queued vehicles, 
resulting from the similar distances and velocities that
they share inside traffic jams.
While the propagation velocity $C$ of traffic jams is given by the
dissolution speed of a jam front, the outflow $Q_{\rm out}$ is related to
the time gap ${\cal T}$ between successive departures from the
traffic jam \cite{chania}. 
\par
The cellular automaton proposed by Nagel and Schreckenberg \cite{NS}
meets the properties (i), (ii), and (vi), and most of the 
other properties can be reproduced by separate variants of it 
\cite{var,krauss,STS}. In particular, the continuous version by
Krau{\ss} {\em et al.} \cite{krauss} and the CA by Barlovich {\em et al.}
\cite{STS} seem to display metastable states. Moreover, the continuous 
version is in good agreement with empirical traffic data \cite{krauss}.
Here, we will present a
``unifying'' cellular automaton which, in a certain
parameter range, reproduces all of the above properties. Remarkably,
the characteristic quantities can be calculated analytically.
Moreover, consistent with {\em macroscopic} traffic models, the 
mechanism of traffic jam formation is deterministic \cite{macro,Hab} 
rather than based on internal fluctuations (``randomization'') \cite{krauss}. 
Being related to the optimal velocity model \cite{MD1}, it originates
from the delayed adaptation to an equilibrium velocity, which, in the
instability region, rapidly decays with growing density \cite{Hab,macro}.
\par
We propose the following simple and fast discrete model that can be
well calibrated to macroscopic traffic data:
To maximize simulation speed, we first choose the time step $\Delta T$
of the temporal update as large as possible. It is limited to 
$\Delta T \approx 1$\,s, since this corresponds to the safe time
headway required for avoiding accidents. The spatial discretization
$\Delta X$ of the road should not be larger than the minimal vehicle
distance $\ell$. 
A fine velocity discretization $\Delta V = \Delta X / \Delta T$ is reached by
taking a fraction $\Delta X = \ell /n$ of $\ell$ ($n \in \{1,2,3\,\dots\}$). 
Velocity steps of the order $5-10$\,km/h require
$\Delta X \le \ell/2$. In the following, the spatial coordinate $X= x\,\Delta X$
and distances $D = d\,\Delta X$
are measured in units of $\Delta X$, time $T=t\,\Delta T$ 
in units of $\Delta T$, and any velocity in units of $\Delta V$. 
\par 
Assume we want to distinguish $A$ different vehicle types 
$a\in\{1,2,\dots,A\}$, e.g. cars and trucks \cite{HelHub}. Then, 
at each time $t\in\{0,1,2,\dots\}$, every site $x\in \{1,2,\dots,l\}$
can be in one of the states $f_t(x) = (a,v)$, where $f_t(x) = (0,0)$ 
corresponds to an empty site, and $f_t(x) = (a,v)$ with $a>0$
represents a vehicle of type $a$ 
with velocity $v \in\{0,1,2,\dots,v_{\rm max}^a\}$.
For safety reasons, the velocity
$v_t$ must be smaller than the distance $d_t$ to the vehicle ahead. 
The states of the cellular 
automaton are updated in parallel
according to the following successive
steps: First, each vehicle is moved by its actual velocity $v_t$ to 
position $x+v_t$, which means $f_{t+1}(x+v_t) = f_t(x)$, if $a>0$
and $v_t > 0$. Then, the states of the previously occupied positions
are reset to $f_{t+1}(x) = (0,0)$. Any other site keeps its 
previous state. Finally, all vehicle velocities 
are modified along with the proposed acceleration rule
\begin{eqnarray}
 v'_{t+1} &=& v_t + \Big\lfloor \lambda_a [ v_a
 (d_{t+1}) - v_t ] \Big\rfloor \, , \\ 
 v_{t+1} &=& v'_{t+1} - \left\{
\begin{array}{l}
1 \mbox{ with probability } p, \mbox{ if } v'_{t+1} > 0 \\
0 \mbox{ otherwise,} 
\end{array} \right. \quad 
\end{eqnarray}
where 
$\lfloor y \rfloor$ is defined by the largest integer $i\le y$.
Therefore, the above equation implies
$v_{t+1} \le \lambda_a v_a(d_{t+1}) + (1-\lambda_a) v_t$,
meaning that the new velocity is a weighted average of the previous
velocity $v_t$ and the optimal velocity $v_a$ of vehicle type $a$, 
or somewhat less. A small value of the model parameter
$\lambda_a \ge 0$ relates to a great inertia of vehicle
motion, whereas a large value $\lambda_a \le 1$ implies a fast 
adaptation to the distance-dependent optimal velocity $v_a(d)$. The
corresponding adaptation time is $\tau_a = \Delta T / \lambda_a$.
If the (back-bumper-to-back-bumper) distance $d_{t+1}$ to the next vehicle 
exceeds a certain finite value
$d_{\rm fin}$, the vehicles do not interact, and $v_a$ is given by 
the maximum velocity $v_{\rm max}^a$ of vehicle type $a$.
For small distances, $v_a$ should be 
determined by the velocity-dependent safe distance $d(v_a) \approx
\ell/\Delta X + v_a $, required to avoid accidents.   
The model parameter (`slowdown probability')
$p$ describes individual velocity fluctuations 
due to delayed acceleration (imperfect driving). Here, we are interested
in the limit $p \to 0$.
\par
In order to compare this discrete model 
with the observed properties of traffic flows, 
it is necessary to investigate
aggregate quantities $\langle h \rangle$. These are 
defined by
\begin{equation}
\langle h \rangle_{x,t}
= \frac{1}{\Delta t} \sum_{t'=t}^{t+\Delta t-1} 
\frac{1}{2\,\Delta x + 1} \sum_{x'=x-\Delta x}^{x+\Delta x} 
 h\big(f_{t'}(x')\big) \, .
\end{equation}
The vehicle density is given by
$\rho(x\,\Delta X,t\,\Delta T)=\langle \Theta(a) \rangle/$\linebreak
$\Delta X$, the traffic flow by
$Q(x\,\Delta X,t\,\Delta T) = \langle v \Theta(a) \rangle/ \Delta T$,
and the average velocity by $V(x\,\Delta X,t\,\Delta T)
= Q(x\,\Delta X,t\,\Delta T) / \rho(x\,\Delta X,t\,\Delta T)$,
where $\Theta(a) = 1$ for $a\ge 0$, otherwise $=0$.
During the simulation runs, the density minima and maxima,
their propagation speed, and the associated traffic flows were
evaluated automatically, as well as the spatially averaged vehicle velocity
and traffic flow on the circular road of length $L=20$\,km. The corresponding
values were averaged over several hours after a sufficiently long
transient period. Averages over the whole street are indicated by an
overbar. 
\par
\begin{figure}[htbp]
\begin{center}\hspace*{-0.5\unitlength}
\epsfig{width=3.8\unitlength, angle=-90, 
      file=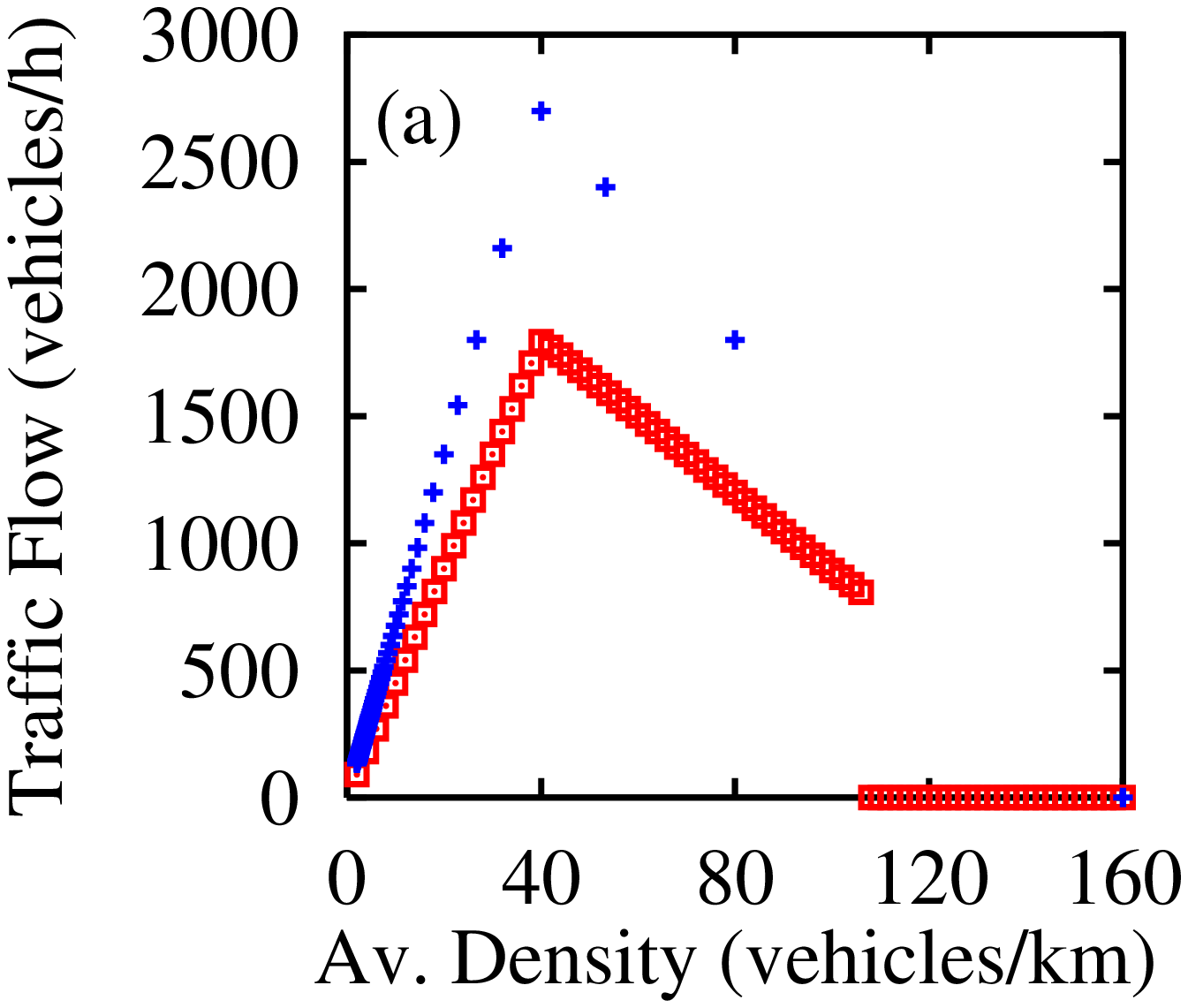} 
\hspace*{-0.1\unitlength}
\epsfig{width=3.8\unitlength, angle=-90, 
      file=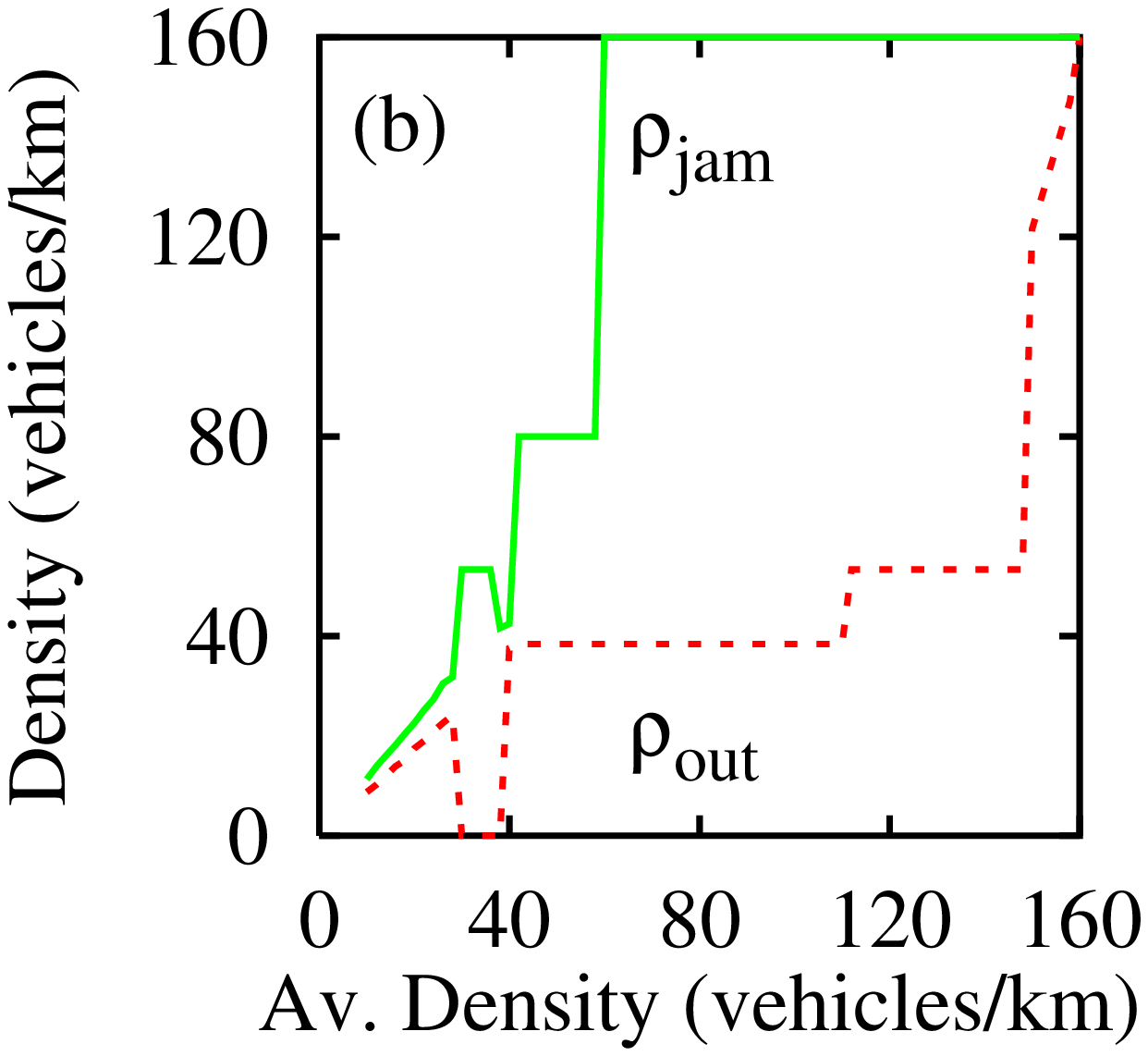} 
\end{center}
\caption[]{Simulation results for $v_1(d)=\min(d-1,3)$,
$\lambda_1 = 0.77$, $\Delta T = 1$\,s, $\Delta X = 6.25$\,m, 
and (a) $p=0.001$, (b) $p=0$, $w_+ = w_- = 200$\,m, and $\Delta \rho
= \rho/2$.
Illustration (a) shows the `optimal flow' $\overline{\rho} 
v_1(1/\overline{\rho})$ (+)
and the resulting average flow $\overline{Q}(\overline{\rho})$ ($\Box$)
as a function of the average density $\overline{\rho}$, (b) the
densities inside (---) and in front of (--~--) traffic jams.\label{FIG1}}
\end{figure}

First, let us discuss the case of one type $a=1$ of ve\-hic\-les. 
Our simulation results can be summarized as follows:
At small average densities $\overline{\rho}$, homogeneous
traffic flow is stable, and the 
spatially averaged velocity $\overline{V}$ is given by the
density-independent value
$( v_1^{\rm max} - \lceil 1/\lambda_1-1 \rceil - p)\Delta V$.
It is zero for high densities [Fig.~\ref{FIG1}(a)], 
since $\lambda_1 v_1(d) > 1$ is necessary for the acceleration
of a vehicle from standstill.
\par
\begin{figure}[htbp]
\begin{center}\hspace*{-0.5\unitlength}
\epsfig{width=3.8\unitlength, angle=-90, 
      file=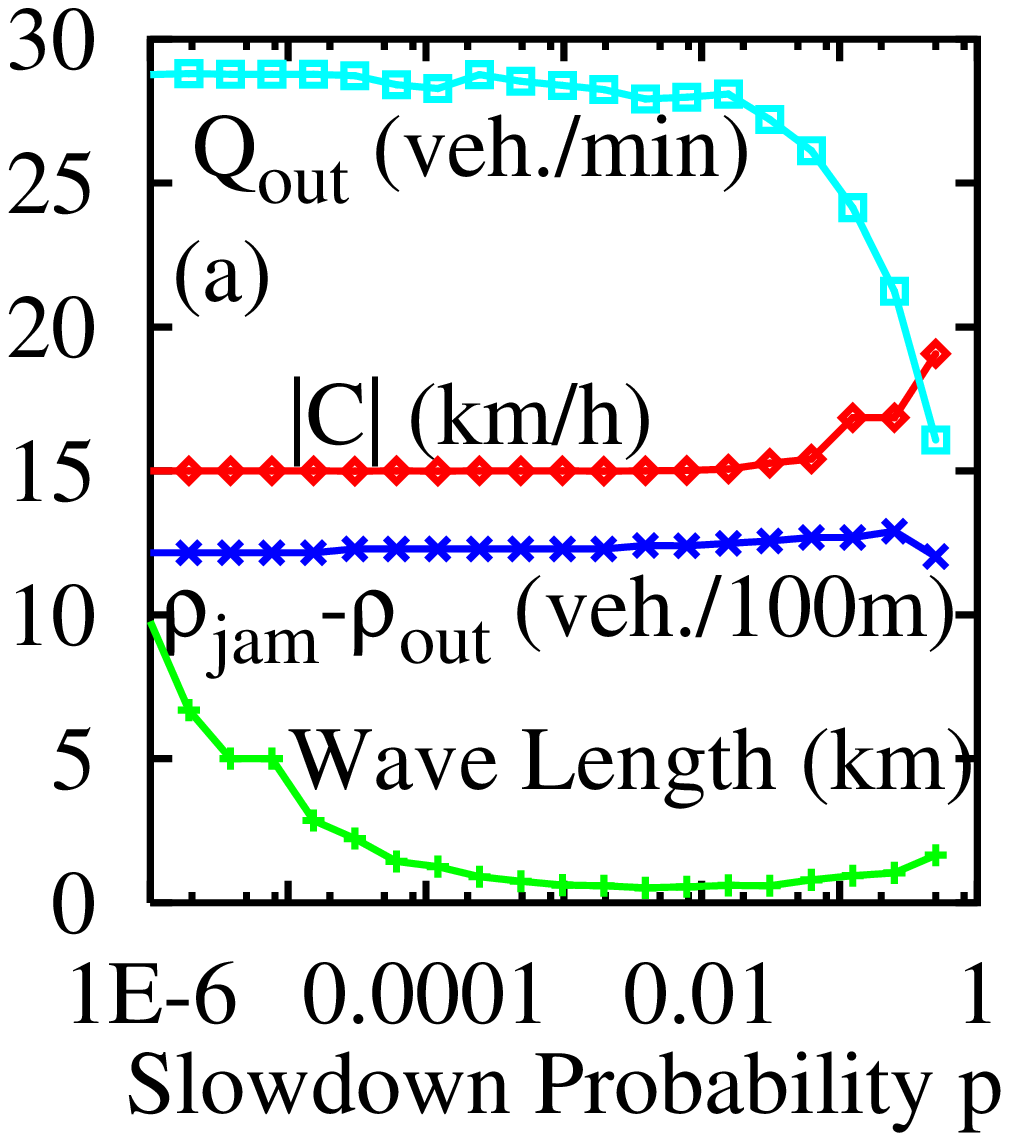} 
\hspace*{-0.5\unitlength}
\epsfig{width=3.8\unitlength, angle=-90, 
      file=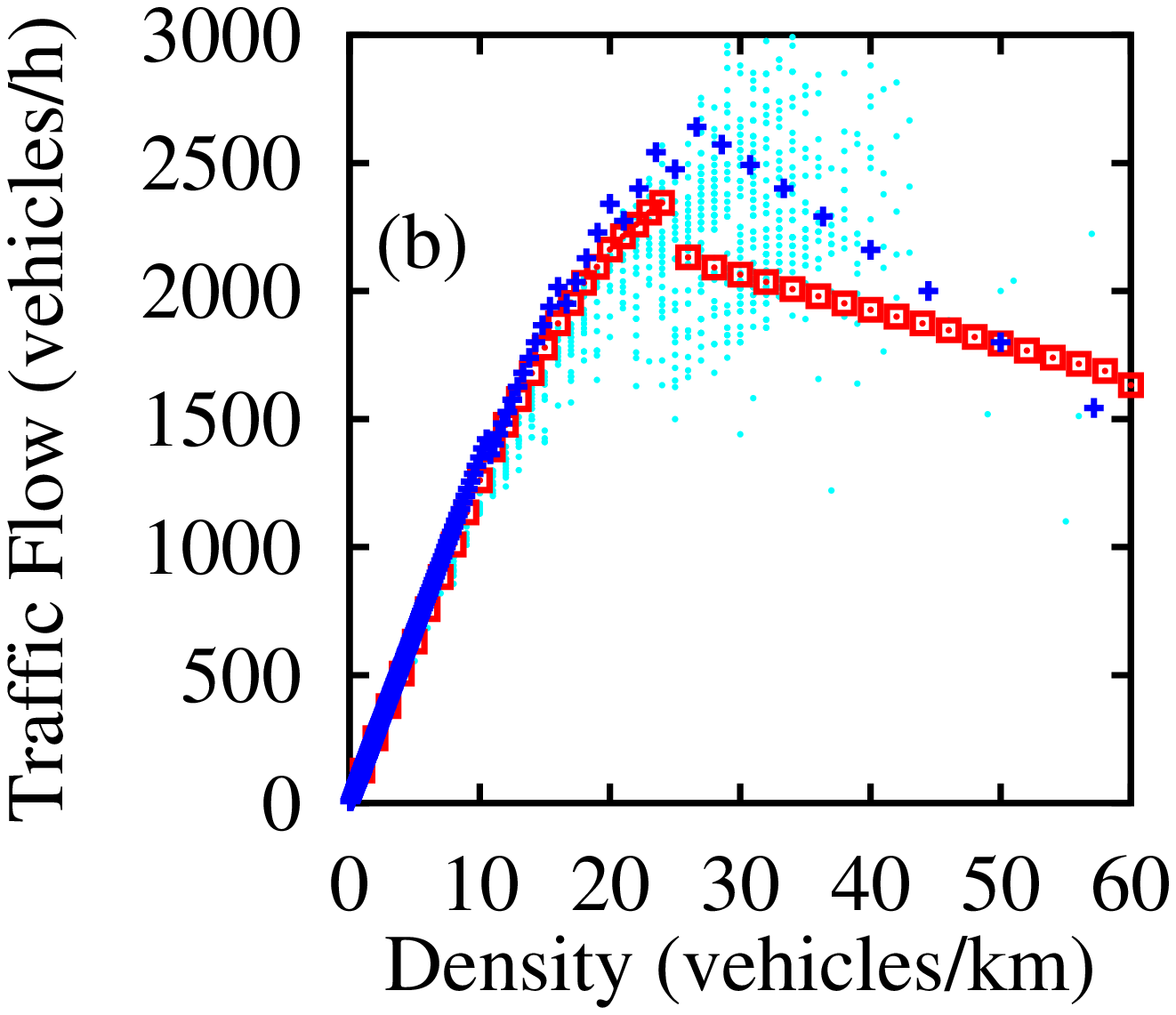} 
\\ \hspace*{-0.8\unitlength}  
\epsfig{width=3.8\unitlength, angle=-90, 
      file=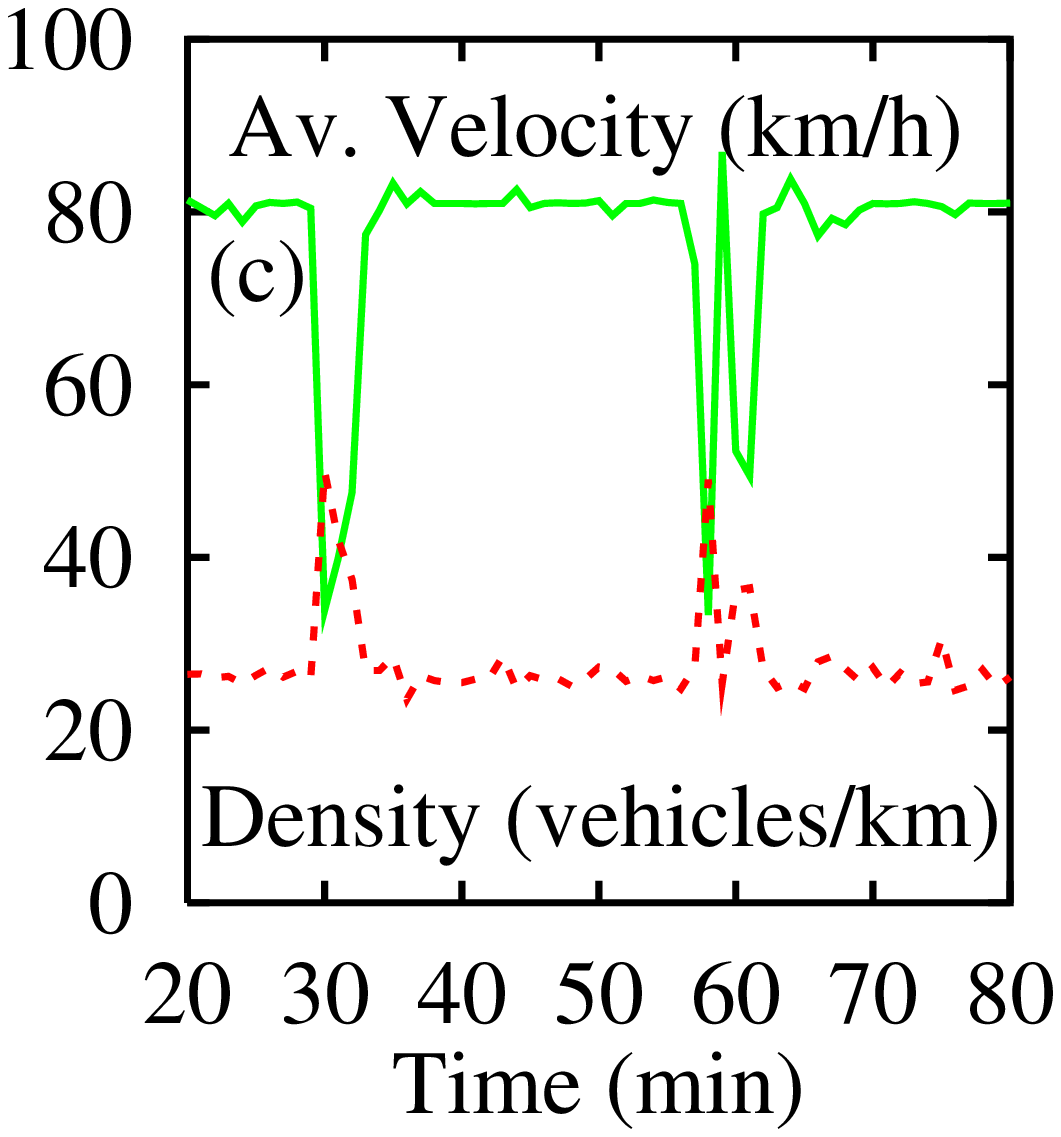} 
\hspace*{0.1\unitlength}
\epsfig{width=3.8\unitlength, angle=-90, 
      file=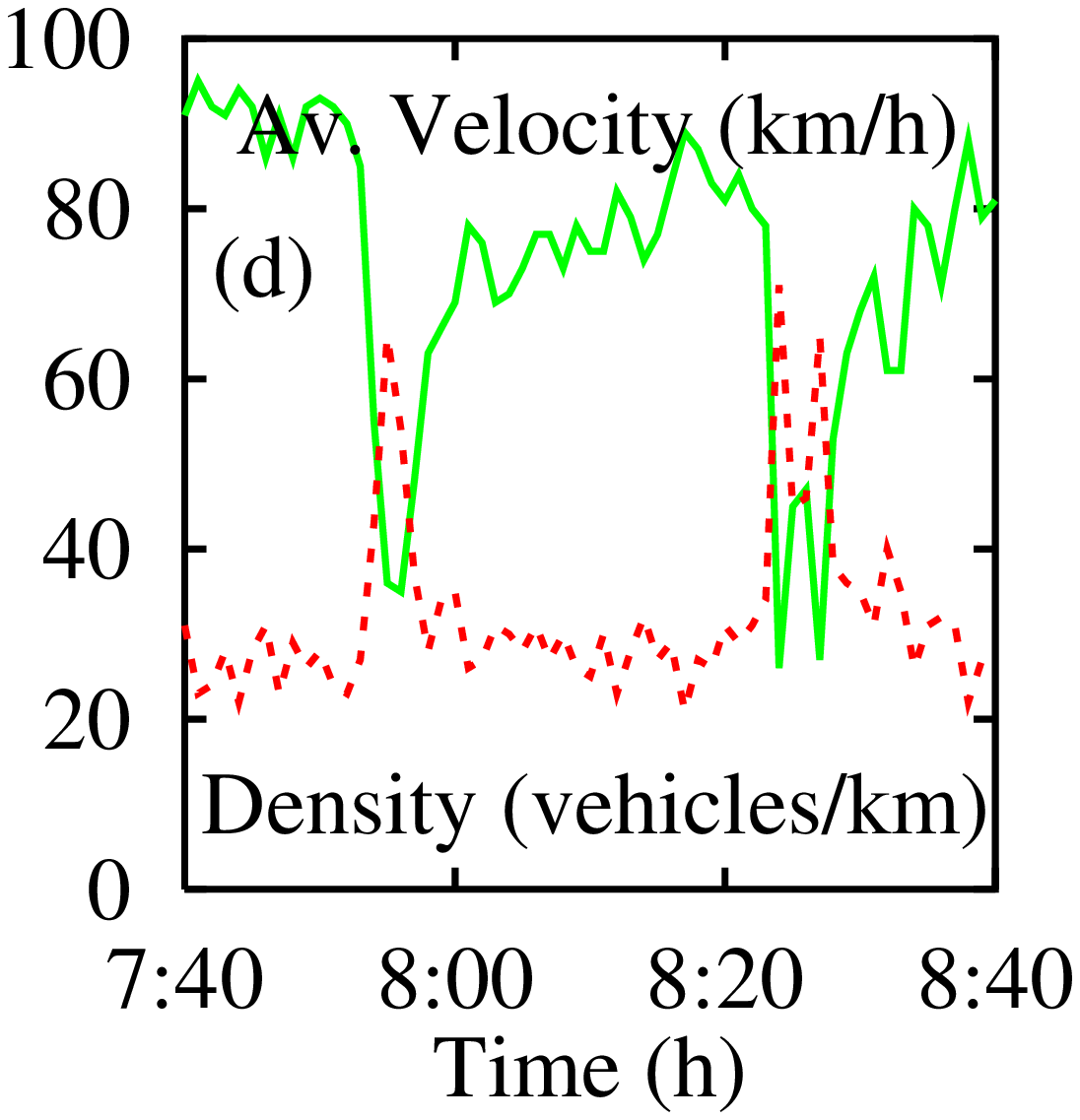} 
\end{center}
\caption[]{(a) Characteristic parameters of traffic flow as a function of
the slowdown probability $p$ for the model parameters 
specified in Fig.~\ref{FIG1}.
(b) Comparison of the model corresponding to $\lambda_a = 0.77$, $p=0.001$,
$\Delta T = 1$\,s, $\Delta X = 2.5$\,m, and the optimal flow relation specified
by '+' with 1-minute-averages of single-vehicle data from the left 
lane of an undisturbed cross section of the Dutch highway A9 ($\cdot$).
Boxes illustrate the simulated average flow resulting in the limit of 
long times.
(c) Simulation of `stop-and-go waves' at an average density of 28\,vehicles/km
for a mixture of 90\% cars and
10\% trucks (the optimal velocities of which are only 70\% of the cars).
The occuring minimal and maximal velocities, the minimal densities, 
and the largely varying time intervals of successive breakdowns of
velocity are in good agreement with the Dutch freeway data 
(1-minute-averages) displayed in (d). The mixture of vehicle types also 
explains the observed fluctuations in the density and average velocity
of vehicles.
\label{FIG2}}
\end{figure}

At medium densities, the resulting 
velocity-density relation is largely dependent on $\lambda_1$: 
Regime I: For $1 \gtrsim \lambda_1 \ge \lambda_{\rm stab} 
(v_{\rm max}^1,p)$ (corresponding to a
quasi-instantaneous adaptation to the
optimal velocity), $\overline{V}(\overline{\rho})$ is close
to the piecewise constant
relation $v_1(1/\overline{\rho})$.
Accordingly, the associated average traffic flow
$\overline{Q}$ is a piecewise linear function of $\overline{\rho}$.
Regime II: In a certain interval $\lambda_{\rm min}(v_{\rm max}^1,p) 
\le \lambda_1 \le \lambda_{\rm max}(v_{\rm max}^1,p)$, 
traffic flow is unstable for a certain range 
$\rho_{\rm out} \le \overline{\rho} \le \rho_{\rm max}$
of medium densities, and $\overline{Q}(\overline{\rho})$
becomes the self-organized linear relation 
\begin{equation}
 \overline{Q}(\overline{\rho}) = \frac{1}{{\cal T}}\left( 1 -
   \frac{\overline{\rho}}{\rho_{\rm jam}} \right)
\label{Qformel} 
\end{equation}
demanded by Kerner \cite{chania} [Figs.~\ref{FIG1}(a) and \ref{FIG2}(b)].
${\cal T}$ denotes the average time gap 
between the acceleration of successive vehicles.
The linear relation (\ref{Qformel}) reflects a mixture of free and 
jammed traffic with characteristic densities $\rho_{\rm out}$ and 
$\rho_{\rm jam}$, respectively, 
where the jammed regions grow with increasing density.
$Q_{\rm out} = \overline{Q}(\rho_{\rm out})$ is the typical outflow
from traffic jams \cite{chania}. The slope
\begin{equation}
 C  = \frac{\partial \overline{Q}}{\partial \overline{\rho}}
 = - \frac{1}{{\cal T} \rho_{\rm jam}} 
\end{equation}
corresponds to their dissolution velocity \cite{chania}. 
The dependence of the
spatially averaged velocity $\overline{V}$ on $\overline{\rho}$ is given
by $\overline{V}(\overline{\rho}) = \overline{Q}(\overline{\rho})/
\overline{\rho}$.
Regime III: 
In a parameter range $\lambda_{\rm max}(v_{\rm max}^1,p) < \lambda_1
< \lambda_{\rm stab}(v_{\rm max}^1,p)$, 
there is still an unstable range of traffic,
but $\overline{Q}(\overline{\rho})$ is {\em piecewise} linear with 
different slopes $C$ and different values of 
$\rho_{\rm jam}$ or $\rho_{\rm out}$ [like in Fig.~\ref{FIG1}(b)], 
where the above relations are separately fulfilled for
each linear piece. This may be understood as crossover behavior between
the cases I and II. Notice that the discretization of vehicle dynamics
implies $1/\rho_{\rm max} = k_1 \,\Delta X$ and
${\cal T} = k_2 \, \Delta T/k_3$ with small integers $k_i$ (see below).
Thus, $C$ is restricted to a few
discrete values $\frac{k_1 k_3}{k_2} \Delta V$. 
Regime IV: For $\lambda_1 v_{\rm max}^1<1$, $\lambda_1$ is so small
that $v'_{t+1} \le v_t$ which, for $p\ne 0$, implies that traffic eventually
comes to rest.  
\par
The characteristic quantities can explicitly be calculated.
Let us show this for the simple but
nontrivial case $p \to 0$ 
and $v_1(d) = \min(d-1,3)$ modeling
city traffic. With $\Delta X = 6.25$\,m we have
$\rho_{\rm jam} = 1/\Delta X = 160$\,vehicles/km. Now, let us assume
that a disturbance has produced
a queue of vehicles with velocities $v=0$ and distances $d=1$ to the
respective vehicles ahead, and a free road in front of the first vehicle.
We can characterize the acceleration behavior of a vehicle by 
the sequence 
\begin{equation}
(v_{t'},d_{t'}) \stackrel{v_{t'}^*}{\to} 
(v_{t'+1},d_{t'+1}) \stackrel{v_{t'+1}^*}{\to} (v_{t'+2},d_{t'+2}) 
\stackrel{v_{t'+2}^*} \to \dots \, ,
\end{equation}
where, for $\lambda = 0.77$, $v_{t+1} = \max(d_t -2,0)$ 
if $(v_t+1) \le d_t\le 4$, and
$d_{t+1} = (d_t + v_{t+1}^* - v_{t+1})$ ($v_{t+1}^*$ being the
velocity of the vehicle ahead). Denoting with $t'$ the time 
when the state $(0,1)$ of the respective vehicle ahead has changed
to another state, we find two alternating sequences:
$(0,1) \to (0,1) \to (0,3) \to (1,4) \to (2,4) \to (2,4) \to \dots$ and
$(0,1)\to (0,2) \to (0,4) \to (2,4) \to (2,4) \to \dots$ That is, 
cars start to accelerate alternatingly every one or two time steps $\Delta T$.
With $\Delta T = 1$\,s, we find an average of ${\cal T} = 1.5$\,s,
which implies the dissolution velocity $C = -15$\,km/h. Moreover,
the resulting density in front of jams is given by the evolving maximal
distance $4\Delta X$, which gives $\rho_{\rm out} = 40$\,vehicles/km 
and, according to Eq. (\ref{Qformel}),
an outflow of $Q_{\rm out} = 1800$\,vehicles/h. 
All this is in total agreement with simulation results 
[Fig.~\ref{FIG1}(a)].
\par
Now, let us investigate the behavior for fixed $\lambda_1 = 0.77$ 
and density $\overline{\rho} = 80$\,vehicles/km, but
various $p$ [Fig.~\ref{FIG2}(a)]. If $p$ is continuously reduced, 
the backward dissolution 
velocity $C$ is decreasing, whereas the outflow $Q_{\rm out}$ 
from traffic jams, and the difference between
the jam density $\rho_{\rm jam}$ and the self-organized density
$\rho_{\rm out}$ downstream of traffic jams are rapidly growing towards
an almost constant value. In particular, the amplitude 
$(\rho_{\rm jam} - \rho_{\rm out})$ of traffic jams is constant over
more than five decades. 
\par
In the deterministic case $p=0$, the formation of
traffic jams requires some initial inhomogeneity.
Let us investigate the response to localized perturbations of the form
$\rho(x,0) = \overline{\rho} + \Delta \rho
\{ \mbox{cosh}^{-2}[(x-L/2)/w_+] - (w_+/w_-)$ $\mbox{cosh}^{-2}
[(x-L/2-w_+ -w_-)/w_-]\}$,
as suggested in Ref. \cite{MD2}. Typically, one observes 
a piecewise linear response like in regime III [Fig.~\ref{FIG1}(b)]. 
However, its dependence on the perturbation amplitude $\Delta \rho$ 
indicates multistability,
i.e. the coexistence of a variety of solutions. These correspond to 
periodic patterns of vehicle updates (which naturally result 
in deterministic systems with a {\em finite} number of states). 
In the presence of noise
($p>0$), only one of them survives [Fig.~\ref{FIG2}(a)], i.e.
most of them are unstable with respect to fluctuations.
This explains the role of randomization for the behavior resulting in
regime II. 
\par
\begin{figure}[htbp]
\begin{center}\hspace*{-0.5\unitlength}
\epsfig{width=3.8\unitlength, angle=-90, 
      file=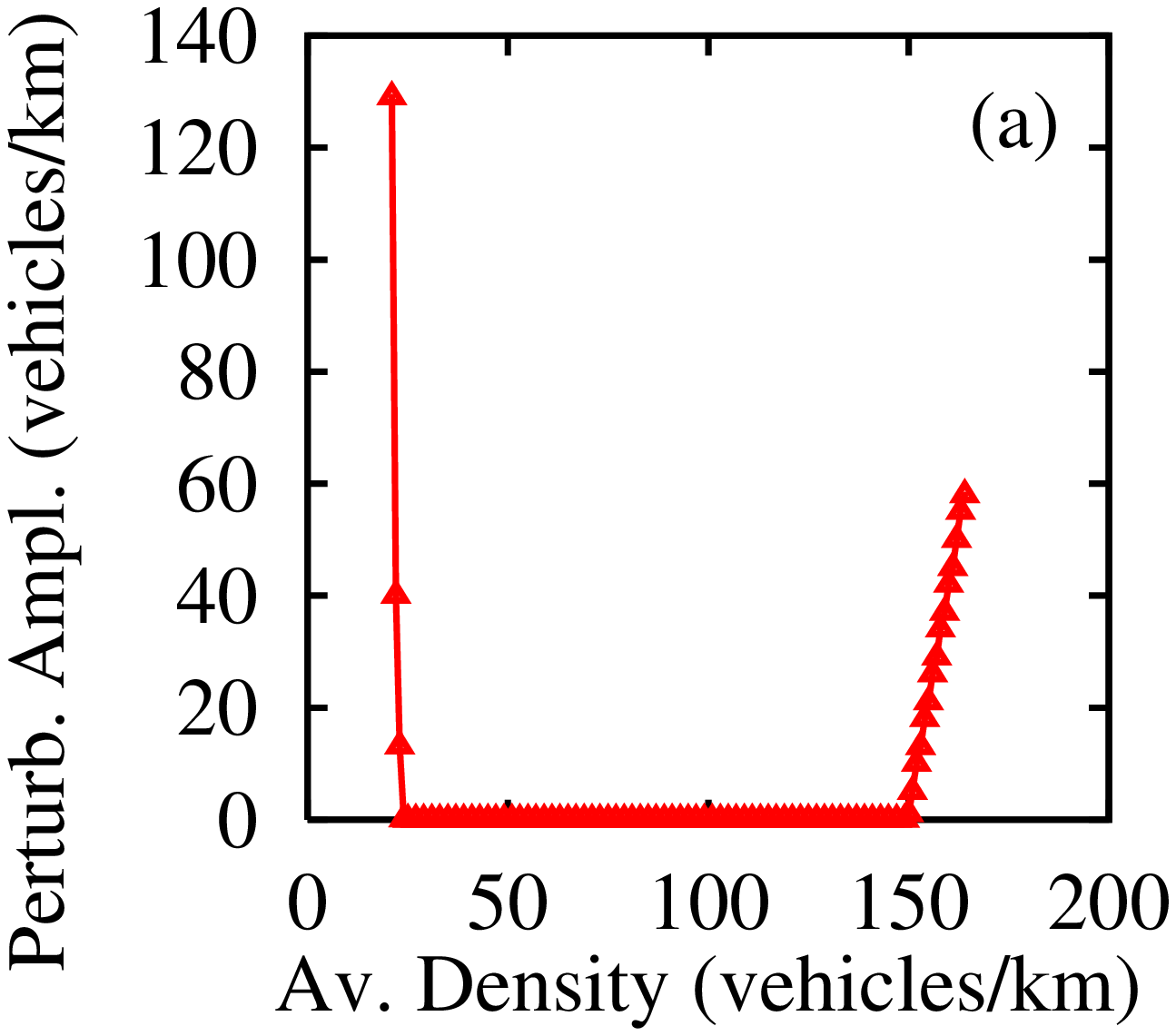} 
\hspace*{-0.5\unitlength}
\epsfig{width=3.8\unitlength, angle=-90, 
      file=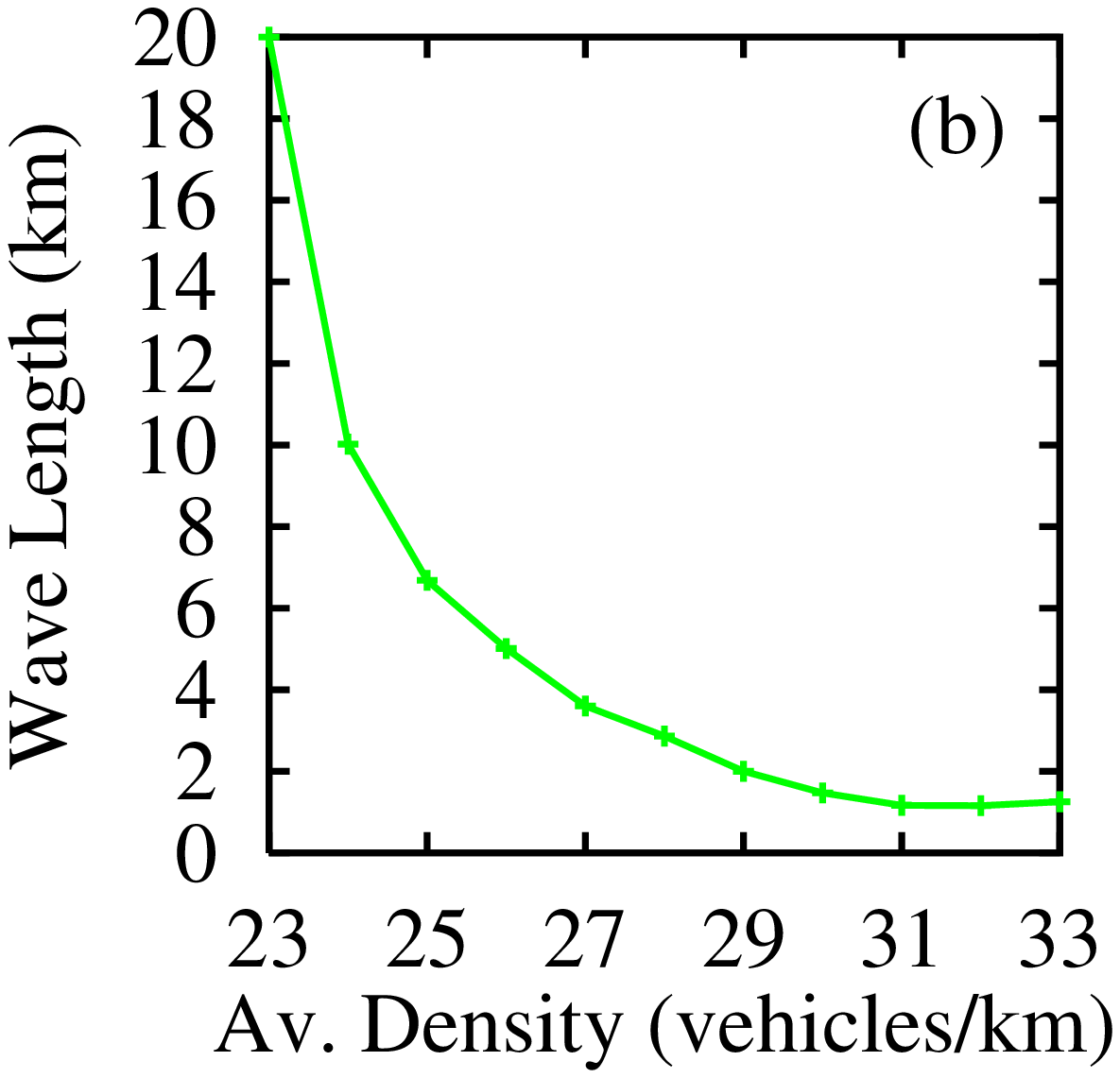} 
\end{center}
\caption[]{(a) Critical amplitudes 
of localized perturbations with $w_+=200$\,m and $w_- = 800$\,m
as a function of average density
for the model specified in Fig.~\ref{FIG2}(b). Whereas
larger perturbations cause the formation of traffic jams, smaller ones
will fade away ($\rho_{\rm c1} = 21$\,vehicles/km,
$\rho_{\rm c2} = 23$\,vehicles/km, $\rho_{\rm c3} = 150$\,vehicles/km,
$\rho_{\rm c4} \ge 164$\,vehicles/km). (b) The average wave length of
emerging stop-and-go waves diverges at $\rho_{\rm c2}$ (checked for
large system sizes). This is why
they are not triggered by fluctuations below this density. 
\label{FIG3}}
\end{figure}

Finally, let us focus on the density-dependent behavior for
fixed $\lambda_1 = 0.77$ and $p > 0$. We find that initial localized
perturbations of the above form, regardless of their amplitude $\Delta \rho$,
are damped out for average densities
$\overline{\rho}$ below some value $\rho_{\rm c1}(v_{\rm max}^1,p)$ and
above some value $\rho_{\rm c4}(v_{\rm max}^1,p)$. In a certain density
range $\rho_{\rm c2}(v_{\rm max}^1,p) < \overline{\rho} <
\rho_{\rm c3}(v_{\rm max}^1,p)$, the perturbation grows 
for any finite amplitude. In the density regimes   
$\rho_{\rm c1}(v_{\rm max}^1,p) \le \overline{\rho} \le
\rho_{\rm c2}(v_{\rm max}^1,p)$ and 
$\rho_{\rm c3}(v_{\rm max}^1,p) \le \overline{\rho} \le
\rho_{\rm c4}(v_{\rm max}^1,p)$, we observe metastability [Fig.~\ref{FIG3}], 
i.e. perturbations with an amplitude $\Delta \rho 
\ge \Delta \rho_{\rm cr}(\overline{\rho})$ will grow, otherwise
they will fade away in the course of time 
(`local cluster effect' \cite{macro,MD2}). Notice that
$\rho_{\rm c2} \approx \rho_{\rm out}$ and $\rho_{\rm c3}
\approx \rho_{\rm max}$.
\par
These findings can be understood
by analogy with the continuous optimal velocity model by
Bando, Sugiyama {\em et al.} \cite{MD1} that results in the limit
$\Delta T \to 0$ and $\Delta V \to 0$. It displays stable traffic
at low and high vehicle densities, unstable
traffic on the condition $\mbox{d}v_1(d)/\mbox{d}d > \lambda_1/2$,
and metastable regimes between the unstable and stable ones \cite{MD1}. 
However, in the {\em continuous} optimal velocity model, the jam density
$\rho_{\rm jam}$ is not independent on how a traffic jam is formed,
since fast cars are more crowded than slow ones, after they had to stop. 
This undesired property can be avoided by a refined model
\cite{gfm} or just by a suitable
discretization, like in the proposed cellular automaton, when operated in
regime II. 
\par
In summary, we have developed a cellular automaton for one-lane
traffic which reproduces many of the empiricially observed features
of traffic flow in a ``unified'' way. In particular, the model
showed the characteristic quantities of traffic flow, which we
managed to calculate analytically. 
By appropriate specification 
of the tabular functions $v_a(d)$ and the parameters $\lambda_a$, $p$, 
$\Delta T$, and $\Delta V$, 
the model can be calibrated to empirical data. 
The most interesting case
is to operate the model in regime II, since this guarantees the
desired properties (iv) and (vi). The  
characteristic quantities like $C$ and $Q_{\rm out}$ are determined by
$\rho_{\rm out}$, $\rho_{\rm jam} = k_1\,\Delta T$, 
and ${\cal T} = k_2\,\Delta T/k_3$. The latter can 
be enforced by a suitable choice of $\Delta X$ and $\Delta T$. 
$v_a(d)$ and $\lambda_a$ determine the maximum velocity,
the approximate velocity-density relation,
the instability region, and the amplitude $(\rho_{\rm jam} - \rho_{\rm out})$
of traffic jams. The metastable regimes and
the difference between the maximal possible traffic flow and the
self-organized outflow $Q_{\rm out}$ from traffic jams grow with
increasing $v_{\rm max}^a$. Finally, $p$ allows to influence the characteristic
`wave length' between successive traffic jams [Fig.~\ref{FIG2}(a)].
Suitable choices are $\Delta T \in [1\,\mbox{s},1.3\,\mbox{s}]$, 
$\lambda_a \approx 0.77$ and $p \le 0.01$. 
The optimal velocity functions $v_a(d)$ were chosen
proportional to relations that were 
determined from traffic data of the Dutch freeway A9. 
The results are in good agreement with macroscopic traffic data
[Fig.~\ref{FIG2}(b)--(d)]. 
\par   
The authors want to thank for financial support by the BMBF (research
project SANDY) and the DFG (Heisenberg scholarship
He 2789/1-1). They are grateful to B. Kerner
for valuable comments, and to H. Taale and 
the Dutch {\it Ministry of Transport,
Public Works and Water Management} for supplying the freeway data. 

\vspace*{-5mm}
\vspace*{-5mm}
\end{document}